\documentclass[prb,twocolumn,showpacs,superscriptaddress]{revtex4}

\usepackage[dvips]{graphicx}
\usepackage{color}
\usepackage{latexsym}
\usepackage{amsmath}
\usepackage{amsthm}
\usepackage{amsfonts}
\usepackage{amssymb}

\usepackage{bm}
\usepackage{bbm}

\newcommand{\rem}[1]{}
\newcommand{\refe}[1]{(\ref{#1})}
\newcommand{\fige}[1]{Fig.~\ref{#1}}
\newcommand{\refE}[1]{Eq.~(\ref{#1})}
\newcommand{\beq}{\begin{equation}}
\newcommand{\eeq}{\end{equation}}
\newcommand{\beqa}{\begin{eqnarray}}
\newcommand{\eeqa}{\end{eqnarray}}
\newcommand{\eps}{\varepsilon}

\begin{document}

\title{Interplay of magneto-elastic and polaronic effects\\
in electronic transport through suspended carbon-nanotube quantum dots}

\author{G.~Rastelli}
\affiliation{Univ. Grenoble 1/CNRS, LPMMC UMR 5493, Maison des Magist\`{e}res, 38042 Grenoble, France}

\author{M.~Houzet}
\affiliation{SPSMS, UMR-E 9001 CEA / UJF-Grenoble 1, INAC, Grenoble, F-38054, France}

\author{L.~Glazman}

\affiliation{Departments of Physics, Yale University, New Haven, Connecticut 06520, USA}

\author{F.~Pistolesi}

\affiliation{Univ. Bordeaux, LOMA, UMR 5798, F-33400 Talence, France}
\affiliation{CNRS, LOMA, UMR 5798, F-33400 Talence, France}

\date{\today}

\begin{abstract}
We investigate the electronic transport through a suspended
carbon-nanotube quantum dot.
In the presence of a magnetic field perpendicular to the 
nanotube and a nearby metallic gate, two forces act on the electrons: 
the Laplace and the electrostatic force.
They both induce coupling between the electrons and 
the mechanical transverse oscillation modes. 
\rem{
We derive the current to all orders in these magneto-elastic and 
polaronic couplings, but 
at temperatures larger than the electron tunneling rate between 
the nanotube and the leads.
In the sequential tunneling regime we find that the effects of the 
electric and magnetic  fields cannot be distinguished apart from the 
removal of the spin degeneracy induced by the latter.
}
We find that the difference between the two mechanisms appears 
in the cotunneling current.
\end{abstract}


\maketitle

\section{Introduction}

Displacements of conducting nano-mechanical systems are conveniently
registered by the electron transport measurements. 
The electron transport is sensitive to the mechanical motion mostly
due to the electrostatic interaction, which allowed 
one to detect, e.g. the motion
of tiny single systems, like C$_{60}$ molecules\cite{park:2000} or 
carbon nanotubes.\cite{sazanova:2004} 
The latter are particularly promising systems for many applications.
High mechanical quality factors of the order of $10^5$ have been reported
for suspended carbon-nanotube oscillators.\cite{huettel:2009} 
At the same time, strong mechanical coupling to electronic transport, both in the 
bending\cite{lassagne:2009,steele:2009} and breathing modes,\cite{leroy:2004,leturcq:2009}
have also been demonstrated.
The effect of the electrostatic coupling 
to the mechanical oscillations (also known as {\it polaronic} coupling) on transport  
has been investigated by several authors from the theoretical point of 
view.\cite{glazman:1988,wingreen:1989,boese:2001,mccarthy:2003,braig:2003,mitra:2004,zazunov:2006,egger:2008}
A suppression of the current at low bias voltage, 
known as Frank-Condon blockade, has been predicted\cite{braig:2003,koch:2005}
and observed\cite{leturcq:2009} in the quantum regime, for high frequency molecular 
oscillators.
Similar effects in the classical limit (low frequency oscillators) have also been  
discussed.\cite{doiron:2006,mozyrsky:2006,pistolesi:2007,pistolesi:2008,weick:2010,weick:2011}

\begin{figure}%[htbp]
\includegraphics[scale=0.22,angle=0.]{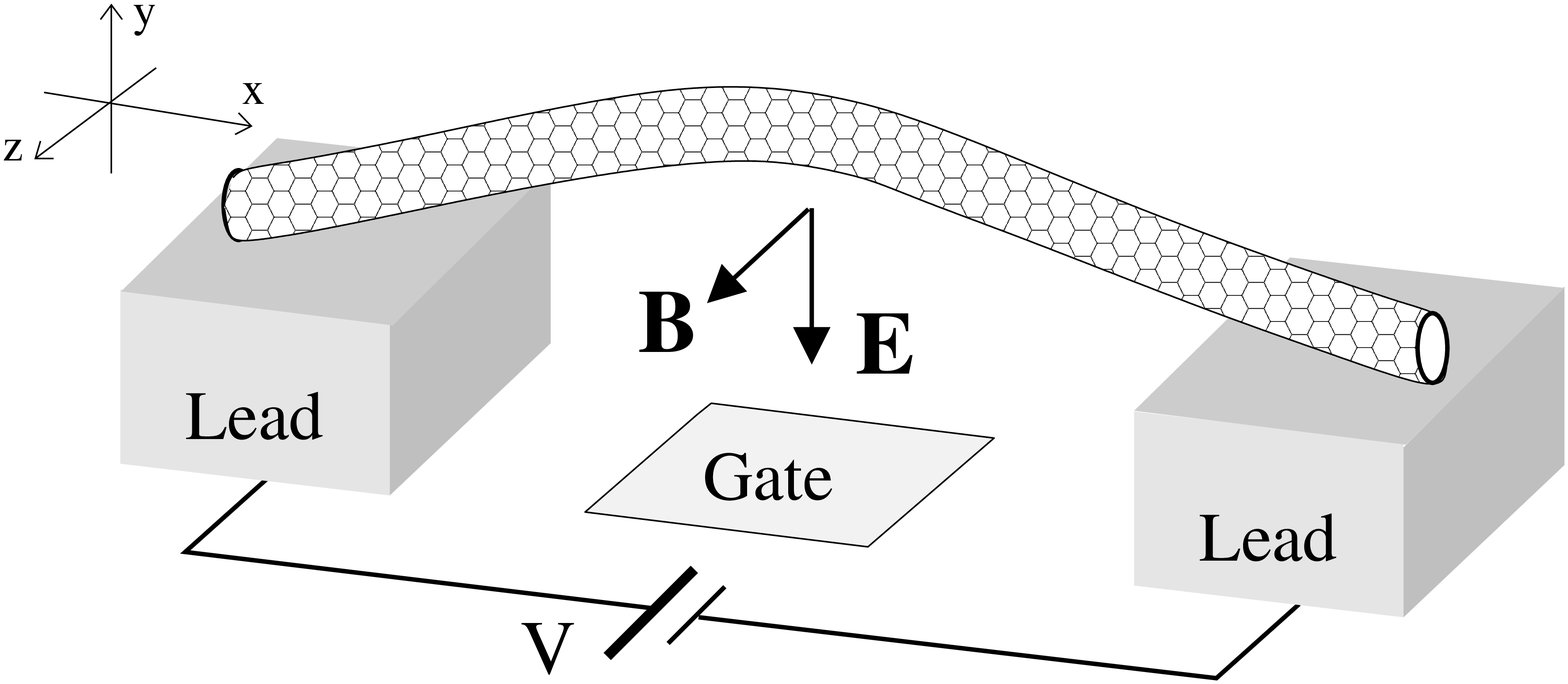}
\caption{Schematic picture of the system:
a voltage biased suspended carbon nanotube
in presence of a transverse magnetic field $\bm{B}$ and of a transverse
electric field $\bm{E}$ created by applying a gate voltage.
}
\label{fig1}
\end{figure}

Much less investigated is the effect of magnetic field, see Fig. \ref{fig1},
on the coupling between electronic transport and 
mechanical oscillations (called {\it magneto-elastic} coupling hereafter).
This coupling is at the origin of the magnetomotive effect, which 
has been used to activate and detect the classical motion of 
micro and nano-mechanical resonators since a long time.
However its manifestation in the quantum regime 
has been addressed only recently for a suspended carbon 
nanotube forming a quantum dot. \cite{shekhter:2006,sonne:2009,rastelli:2010,skorobagatko:2011} 
A particularly appealing interpretation of the magnetoresistance 
in this system has been put forward in Ref. \onlinecite{shekhter:2006}:
In the tunneling limit the electrons
traversing the nanotube interfere over the different paths
generated by its zero point oscillations.
The reduction of the current can then be seen as a joint manifestation 
of the Aharonov-Bohm effect and the quantum fluctuations of the 
mechanical degree of freedom. 
In the resonant case the magnetoresistance has been calculated 
as an expansion up to the second order in the small electromechanical 
coupling.
It has been found that multiple interferences 
increase the size of the effect parametrically.\cite{rastelli:2010}

In any realistic device both the electrostatic and magnetic effects 
are present.
The aim of the present paper is to discuss the interplay
of these two interactions. 
Specifically the system considered is shown schematically in \fige{fig1}.
%
%
%%%%%%%%%%%%%%%%%%%%%%%%%%%%%%
%
%				Figure 1
%
%%%%%%%%%%%%%%%%%%%%%%%%%%%%%%
%
%%%%%%%%%%%%%%%%%%%%%%%%%%%%%%%
%
It consists in a suspended carbon nanotube forming a quantum dot contacted 
electrically to two external leads.
We focus on the coupling of the fundamental bending mode of the nanotube 
to the electronic transport in presence of both a strong magnetic field 
perpendicular to the nanotube and a metallic gate.
They induce a magneto-elastic and a polaronic coupling, respectively.
We consider a high temperature regime when electron tunneling events between the leads can be described in terms of rate equations.\cite{koch:2006,pistolesi:2009} 
On the other hand,  the mechanical mode 
is in the low-temperature limit and, for simplicity, we will further assume that its coupling 
to the environment is sufficiently strong to keep the oscillator 
in equilibrium.

The theory that is formulated in the reminder of the paper allows  
to explore the purely magnetic problem outside the 
limits of previous works,\cite{shekhter:2006,sonne:2009,rastelli:2010}
by investigating transport at bias voltages larger than the oscillator's energy quantum.
The same approach is then used to 
study the interplay between the magneto-elastic and 
polaronic couplings.
In the sequential tunneling regime, we find that the current 
depends on a single coupling constant which is the root mean 
square of the polaronic and magneto-elastic 
coupling constants, respectively.
The presence of a magnetic field at this order thus only renormalizes  
the effective interaction and introduces a Zeeman splitting of the 
electronic levels.

The difference between the two interactions appears 
in the cotunneling regime (when electrons tunnel directly between the 
leads through virtual states in the dot).
When the single energy level in the dot is close or in the 
conducting window between the left and right chemical 
potential of the leads (resonant transport)  cotunneling events 
induce a small correction to the current.
However, they can dominate transport away from resonance.
It is thus in that regime that the difference between the 
polaronic and magneto-elastic couplings becomes clear-cut
and could be detected experimentally.

The paper is organized as follows. 
Section \ref{model} is dedicated to the microscopic 
derivation of the model Hamiltonian used in the reminder 
of the paper. 
Section \ref{formalism} introduces the formalism used to 
tackle the problem.
In Section \ref{sequential} we discuss the behavior of the current 
as obtained in the sequential tunneling approximation and 
the consequences of the Zeeman splitting.
Section \ref{cotunneling} considers the contribution of the cotunneling 
events, discussing in particular the difference between the polaronic and
magneto-elastic couplings.
Section \ref{conclusions} gives our conclusions.
Part of the material is presented in the appendices.

%%%%%%%%%%%%%%%%%%%%%%%%%%%%%
%
%			Section II  MODEL
%
%%%%%%%%%%%%%%%%%%%%%%%%%%%%%

\section{Model Hamiltonian}
\label{model}

We derive the Hamiltonian for a suspended nanowire weakly electronically coupled to leads
in presence of polaronic and magneto-elastic effects in Secs. \ref{magneto}-\ref{combined},
see Eqs. \refe{eq:H_tot_1}-\refe{eq:Htun_L}. We estimate the amplitudes of these
effects for a quantum dot formed with a suspended carbon nanotube in Sec. \ref{discussion}. 

\subsection{Magneto-elastic coupling}

\label{magneto}

In this subsection we present a detailed microscopic derivation 
of the magneto-elastic interaction. 
It extends the model proposed in the previous works.\cite{shekhter:2006,rastelli:2010}

\subsubsection{Electron in a vibrating nanowire}

The Hamiltonian that describes an electron propagating in a
ballistic, suspended nanowire in presence of a transverse magnetic field
(units with $\hbar=k_B=1$ are used throughout the paper) is
\beq
\label{eq:Hamiltonian}
{\cal H}
=
\frac{1}{2m}\left(\bm{p}+e \bm{A}(\bm{r})\right)^2
+V(x,y-u(x),z)+H_\mathrm{string}.
\eeq
Here, $\bm{p}=(p_x,p_y,p_z)$ and $\bm{r}=(x,y,z)$ are the momentum and position
of the electron of mass $m$ and charge $-e$,  $\bm{A}=(-By,0,0)$
is the vector potential associated with the magnetic field $\bm{B}=(0,0,B)$.
The potential
\beq
V(x,y-u(x),z)=V_\mathrm{b}(x)+V_\mathrm{conf}(y-u(x),z)
\eeq
contains two terms. The first one describes the tunnel barriers between the
nanowire and the leads,
\beq
\label{eq:Vb}
V_\mathrm{b}(x)=V_l\delta(x)+V_r\delta(x-L),
\eeq
where $L$ is the nanowire's length. The second one
is the confining potential $V_\mathrm{conf}$ inside the nanowire, it depends
on the deflection $u(x)$ of the nanowire along the $y$-direction.

The elastic bending of the nanowire in the $(x,y)$-plane is described within the
elastic string theory\cite{landau7} with the term
\beq
H_\mathrm{string}
=
\int_0^L ds
\left[
\frac{\pi^2}{2\rho}
+\frac{EI}{2}
\left( \frac{\partial^2u}{\partial s^2}\right)^2
\right],
\label{eq:H-string}
\eeq
where $\rho$ is the mass per unit length of the string, $E$, its elastic Young's modulus,
and $I$, its area momentum of inertia, the deflection field and momentum density operators
of the string are conjugated $[u(s),\pi(s')]=i\delta(s-s')$.
For a doubly clamped nanowire, the boundary conditions are $u(s)=\partial_su(s)=0$ at $s=0,L$.

The classical mechanical eigenmodes of the nanowire are characterized by a frequency
$\omega_n$ -- the lowest mode has frequency
$\omega_0\approx 22.4(EI/\rho L^4)^{1/2}$ --
and displacement field $f_n(s)$ (integer $n\geq 0$) with normalization
$(1/L) \int_0^Lds f_n(s)^2=1$. Decomposing the deflection field operator onto these modes,
\beq
\label{eq:modes}
u(s)=\sum_n X_n f_n(s),
\eeq
the Hamiltonian \refe{eq:H-string} then becomes:
\beq
H_\mathrm{string}
=
\sum_n\left(
\frac{P_n^2}{2M}+\frac{M\omega_n^2X_n^2}{2}
\right),
\eeq
where $X_n$ and $P_n$ are the effective positions and momenta of the modes,
with conjugation relation $[X_n,P_m]=i\delta_{nm}$,
and $M=\rho L$ is the mass of the nanowire.

We can reasonably  expect that the confining potential
enforces the replacement $y\rightarrow u(x)$
in \refE{eq:Hamiltonian}, assuming
that the nanowire stands along a line with $y=0$ in classical equilibrium.
The one-dimensional propagation of the electron
and its coupling to the vibrations would then be described by
\beq
\label{eq:Hamiltonian2}
{\cal H}
\approx
\frac{1}{2m}[p_x-eBu(x)]^2+V_\mathrm{b}(x)
+H_\mathrm{string}.
\eeq
A unitary transformation ${\cal H}\rightarrow e^{i{\cal S}
}{\cal H}e^{-i{\cal S}}$ where
${\cal S}=-eB U(x)$, with $\partial_x U(x)=u(x)$ would then yield:
\beq
{\cal H}
={\cal H}_0+\sum_n\left(
\frac{[P_n+eBF_n(x)]^2}{2M}+\frac{M\omega_n^2X_n^2}{2}
\right),
\label{eq:Hamiltonian3}
\eeq
where ${\cal H}_0=p_x^2/(2m)+ V_\mathrm{b}(x)$ and 
$F_n(x)=\int^x ds f_n(s)$, which will be the starting point of the next section.

Appendix \ref{app:derivation_H1d} is devoted to a rigorous derivation
of \refE{eq:Hamiltonian3} starting from \refE{eq:Hamiltonian} and taking into account the
transverse spreading of the electron's wavefunction in the nanowire.

\subsubsection{Single level regime}

Here we simplify further the Hamiltonian \refe{eq:Hamiltonian3}
in the regime of a  single resonant electronic level in the nanowire.

Indeed, the term ${\cal H}_0$ in \refE{eq:Hamiltonian3}  describes an electron propagating
through a double-barrier system. For strong tunnel barriers, quasi-localized states
are formed in the central region.
Due to the coupling with the leads in the outer regions,
they decay with a lifetime $1/\Gamma$.
Assuming a large energy level spacing $\Delta\gg \Gamma$
between the quasi-localized states, we can focus on a single level
with energy $\eps_d$ in the central part of the nanowire,
provided that the chemical potential $\mu$ in the leads is also close to the single level energy,
$|\mu-\eps_d|\ll\Delta$.

In this regime, the single-particle Hamiltonian
${\cal H}_0$ entering \refE{eq:Hamiltonian3}
can be written equivalently
\beq
{\cal H}_0={\cal H}_L+{\cal H}_d+{\cal H}_T,
\eeq
where
\beqa
{\cal H}_L&=& 
\sum_{k,\nu}\eps_kc^\dagger_{k\nu}c_{k\nu}
,\\
%\eeqa
%\beqa
{\cal H}_d&=&\eps_d d^\dagger d,
\\%\eeqa
%and
%\beqa
{\cal H}_T&=&\sum_{k,\nu} (t_\nu c^\dagger_{k\nu} d +\mathrm{h.c.})
\eeqa
describe the leads, the dot, and their tunnel coupling, respectively.
Here, $c_{k\nu}$ ($\nu=l,r$) are the annihilation operators for the
electron propagating states in the left ($l$) and right ($r$) leads, labeled with
momenta $k$, and energies $\eps_k=k^2/2m$;
$d$ is the annihilation operator in a single level state of the nanotube, with energy
$\eps_d=\pi^2p^2/(2mL^2)$ ($p$ is a strictly positive integer).
The wavefunction associated with this
level is $\sqrt{2/L}\sin(\pi p x/L)$.
The tunnel coupling between this state and the continuum of states in the leads
is accounted with the tunnel matrix elements
$t_\nu$. Each  lead contributes to the single-level energy broadening
$\Gamma_\nu=2\pi N_\nu |t_\nu|^2$, respectively, where
$N_\nu$ are the densities of states (per spin) at the Fermi level in the leads 
($\Gamma=\Gamma_l+\Gamma_r$ is the total braodening).
They can be related with the parameters in \refE{eq:Vb} through
the relations $\Gamma_\nu=2\pi^{-1}\Delta(\mu/V_\nu k_F)^2$
at $\nu=l,r$,
where $\Delta\approx\pi^2 p/mL^2$ (at large $p$) and
$k_F$ is the Fermi wavevector.

We use the basis of electron states defined above to write the resolution of
the identity (for the single-particle problem considered here) as $\mathbbm{1}=n_l+n_r+n_d$, where
$n_\nu=\sum_kc^\dagger_{k\nu}c_{k\nu}$ and $n_d= d^\dagger d$  count
whether the electron is in the left lead, right lead, or the dot,
respectively. Inserting it on the right and left sides of
\refE{eq:Hamiltonian3} -
and also assuming  that $\Delta\gg \omega_n$ for the relevant oscillator's eigenmodes, 
we find that the Hamiltonian coupling the states 
with energies close to the Fermi level reads
\beq
 {\cal H}={\cal H}_0
+\sum_n(
M\omega_n^2X_n^2/2
+K_{nl}n_l+K_{nr}n_r+K_{nd}n_d
),
\eeq
with
\begin{subequations}
\label{eq:K}
\beqa
K_{nl}
&=&
\frac{[P_n-eBF_{n}(0)]^2}{2M},
\\
K_{nr}
&=&
\frac{[P_n-eBF_{n}(L)]^2}{2M},
\\
K_{nd}
&=&
\frac{[P_n-eB\langle F_{n}\rangle]^2}{2M}
+
\frac{e^2B^2}{2M}
(\langle F_n^2\rangle-\langle F_n\rangle^2),
\nonumber \\
\eeqa
\end{subequations}
and
\beq
\langle F_n^\alpha \rangle
=
\frac{2}{L}\int_0^Ldx\sin^2\left(\frac{\pi p x}{L}\right) F_n^\alpha(x).
\eeq
for $\alpha=1,2$.

In order to eliminate the $B$-dependent terms in Eqs. \refe{eq:K}, 
we perform another unitary transformation ${\cal H}\rightarrow e^{i{\cal S}'} {\cal H} e^{-i{\cal S} '}$, 
with
\beq
{\cal S}'=-eB\sum_n\left[n_l F_n(0)+n_rF_n(L)+\langle F_n\rangle n_d\right]X_n,
\eeq
 and we find
\beqa
\label{eq:Hamiltonian60}
{\cal H}&=&
H_{\mathrm{string}}
+
{\cal H}_L
+
\tilde{\cal H}_d
+
\tilde {\cal H}_T.
\eeqa
Here,
\beq
\tilde {\cal H}_d=\tilde \eps_d d^\dagger d,
\eeq
where 
\beq
\tilde \eps_d=\eps_d+\frac{e^2B^2}{2M}
(\langle F_n^2\rangle-\langle F_n\rangle^2)
\eeq
includes a diamagnetic shift for the single level,
and
\beqa
\label{eq:tildeHT}
\tilde {\cal H}_T
=\sum_{k} \{t_l\exp( {ieB\sum_nX_n[\langle F_n\rangle-F(0)]} )c^\dagger_{kl} d +\mathrm{h.c.}\}
\nonumber \\
+\sum_k \{t_r \exp({ieB\sum_nX_n[\langle F_n\rangle-F(L)]} ) c^\dagger_{kr} d+\mathrm{h.c.}\}.
\nonumber\\
\label{eq:Hamiltonian5}
\eeqa

Introducing the bosonic annihilation operators $b_n$ for the oscillator's modes (phonons),
with $X_n=X_{n0}(b_n+b_n^\dagger)$
($X_{n0}=1/\sqrt{2M\omega_n}$
is the amplitude of zero-point motion in the eigenmode $n$),
we write 
\beq
\label{eq:Hstring}
H_{\mathrm{string}}=\sum_n\omega_nb^\dagger_nb_n.
\eeq
We also notice that the eigenmodes of the doubly clamped nanotube have a definite parity and, thus,
$F_n(L)-\langle F_n\rangle=(-)^n [\langle F_n\rangle-F_n(0)]$.
We then rewrite \refE{eq:tildeHT} as
\beqa
\tilde {\cal H}_T&=&
\sum_{k\nu} \{t_\nu\exp[i\sum_n(b_n+b^\dagger_n)\phi_{\nu n}] c^\dagger_{k\nu} d +\mathrm{h.c.}\}
\label{eq:Hamiltonian6}
\eeqa
where $\phi_{ln}=(-)^{n+1}\phi_{rn}=\phi_n$ and
\beq
\label{eq:flux}
\phi_n=\frac{4\pi BX_{n0}L}{\Phi_0}
\int_0^L \frac{dx}{L}
\sin\left(\frac{\pi p x}{L}\right)^2
\int_0^x\frac{ds}{L}f_n(s)
\eeq
is the effective magnetic flux, in units of the flux quantum $\Phi_0=2\pi/e$, that characterizes the
coupling between the electron and the eigenmode $n$ of the oscillator.
Integration by parts allows simplifying \refE{eq:flux} into:
\beq
\label{eq:flux2}
\phi_{n}=\frac{\pi BX_{n0}L}{\Phi_0}
\int_0^L \frac{dx}{L}
f_{n}(x)
\eeq
at even $n$, while a similar formula with an extra factor $1-2x/L+\sin(2\pi p x/L)/(\pi p)$ 
in the integrand holds at odd $n$.

Equation \refe{eq:Hamiltonian60} together with \refe{eq:Hamiltonian6} is the final result of this section.
It generalizes the Hamiltonian derived in Ref. \onlinecite{shekhter:2006}
that only considered the lowest eigenmode.
[Note that different conditions
are obtained to justify its validity (see Appendix \ref{app:derivation_H1d}) 
and that the phase \refe{eq:flux2} is smaller by a factor $2\pi$ here.]
The phase \refe{eq:flux} was interpreted there
as an Aharonov-Bohm phase picked by the electrons when they
cross the tunnel barriers, and which depends on
the deflection of the nanowire.

\subsection{Polaron coupling}
\label{sec:polaron}

In this subsection we generalize the model to include
the electrostatic coupling and 
the presence of the spin degree of freedom.
We begin by adding an electrostatic contribution to the Hamiltonian \refe{eq:Hamiltonian6} \cite{sapmaz:2003}
\beq
	H_C = 
	{e^2 N^2 \over 2 C_g[u(s)] }-eV_g N
	\,,
	\label{Hel}
\eeq
where  $C_g$ is the  
gate capacitance (we assume that the lead capacitances 
are negligible with respect to the gate capacitance), 
$V_g$ is the gate voltage, and $N$ is the operator for the total number
of electrons on the dot.
Since we assume that only a single level contributes to transport,
it is possible to write $N=N_0+n_d$, where $N_0\approx C_g V_g/e$ is the number 
of the excess electrons in the filled levels, and $n_d$ is the number of 
electrons in the relevant level.
For small displacement of the oscillator we can express $C_g[u(s)]$ 
in terms of $X_n$: 
\beq
\label{eq:expansion}
	C_g[u(s)]=C_g[0]\left(1-\sum_n a_n  X_n \right)
	\,,
\eeq
where 
\beq
a_n=-\int_0^L\frac{ds}{L} 
\frac 1{C_g[0]} \frac{\partial C_g}{\partial u}[0]f_n(s) .
\eeq
Substituting the expansion \refe{eq:expansion} into \refe{Hel} one obtains (apart from a constant term):
\beq
	H_C = {U_\infty \over 2} \sum_n a_n X_n
	+U_\infty n_{d\uparrow} n_{d\downarrow} 
	+ (\delta \varepsilon-eV_g) \, n_d
	+ H_{\mathrm{pol}} 
	\,,
	\label{Hel2}
\eeq
where $n_{d\sigma}=d_\sigma^\dagger d_\sigma$, 
$n_d=n_{d\uparrow}+n_{d\downarrow}$,
$U_\infty=e^2 /C_g[0]$,
$\delta \varepsilon=U_\infty (2 N_0+1)/2$,
and
\beq
	H_{\mathrm{pol}} = \sum_n \lambda_n \omega_n(b_n+b_n^\dag) n_d
	\,,
	\label{Hpol}
\eeq
with
$\lambda_n=a_n \delta \varepsilon X_{n0} /\omega_n$.
The first term of \refE{Hel2} gives a shift in the equilibrium position
of the oscillator and can be discarded.
The second term gives the intradot Coulomb repulsion.
We will assume that $U_\infty$ is the largest energy scale of the 
problem, allowing to neglect the contribution 
of the double occupancy state [the term proportional
to  $U_\infty$ times $X_n$ can be neglected for the same reason
and it is not shown in \refE{Hel2}].
The third term can be included in the definition of 
$\tilde\varepsilon_d$ since it gives only a renormalization
of the energy of the level.
The last term, defined in \refE{Hpol}, is the seeked polaronic 
coupling, which models the interaction between the oscillation modes and the 
charge fluctuations in the dot.

The expression for the polaronic coupling constant can be calculated by 
using a model of a distributed capacitance:\cite{sapmaz:2003,flensberg:2006} 
\begin{equation}
C_g[u(s)] 
=
\int^L_0 \!\!\!\!ds \,\,  \frac{2\pi\epsilon_0}{\mbox{arccosh}\left([ (h-u(s)]/ R_{\perp}  \right)} 
\label{Cg}
\, ,
\end{equation}
where $h$ is the distance of the nanotube from the substrate,  
$R_{\perp}$ its radius, and $\epsilon_0$ is the vacuum permittivity.
Substituting \refE{Cg} in the definition of $\lambda_n$ 
and calculating the integral for $u=0$ and $R_\perp \ll h$
one obtains:
\beq
\label{eq:polaronic_coupling}
	\lambda_n \approx {e C_g[0] |V_g| \over 2 \pi \epsilon_0 h L } {X_{n0} \over \omega_n}
	\int_0^L {ds\over L} f_n(s),
\eeq
for $C_g[0]V_g \gg e$.

Including the electrostatic coupling and the spin degrees of freedom 
to the Hamiltonian \refe{eq:Hamiltonian6} we obtain:
\begin{equation}
\label{eq:H_tot}
H = H_\mathrm{string} + H_L + H_d +H _T 
	\, ,
\end{equation}
where the term $H_\mathrm{string}$ 
remains unchanged, see \refE{eq:Hstring}.
The inclusion of the spin and  the chemical potentials gives instead 
\beq
H_L
=
\sum_{\nu,k,\sigma}
\xi^{\vphantom{\dagger}}_{\nu k}
c^\dagger_{\nu k\sigma}
c^{\vphantom{\dagger}}_{\nu k\sigma} \, ,
\eeq
where $c_{\nu k \sigma}$ is the annihilation operator
for a state with momentum $k$ and spin $\sigma=\uparrow,\downarrow$
in lead $\nu=l,r$, $\xi_{\nu k}=\eps_{k}-\mu_\nu$,
and the difference of the chemical potentials $\mu_l-\mu_r=eV$ 
is related to the bias voltage $V$ of the junction. 
The  dot local term modified by the presence of the magnetic
field and  the electrostatic interaction reads:
\beq
H_{d}
=
\sum_\sigma \eps_{d\sigma}^{\vphantom{\dagger}} d^\dagger_\sigma d_\sigma^{\vphantom{\dagger}}
+U_\infty n_{d\uparrow} n_{d\downarrow}
+H_{\mathrm{pol}}
\, .
\label{eq:Hdot}
\eeq
Here 
$\eps_{d\sigma}=\tilde \eps_{d}+\delta \varepsilon +\sigma \mu_B B$, 
($\sigma=+$ for $\uparrow$, $\sigma=-$ for $\downarrow$) and  $\mu_B$ is the Bohr magneton. 
Note that we included the Zeeman splitting in the spectrum of the 
dot, but we don't need to include it in the lead spectrum, since 
the density of states of metallic leads can be considered as constant 
at the Fermi energy.
The last term describes tunneling and generalizes the 
one entering \refE{eq:Hamiltonian6} to take into account the spin:
\beq
H_T=
	\sum_{\nu, k,\sigma}\{
t_\nu
\exp[i\sum_n(b_n+b^\dagger_n)\phi_{\nu n}] c^\dagger_{\nu k\sigma} d_\sigma^{\vphantom{\dagger}}
+\mathrm{h.c.} \}
\, .
\label{eq:HamiltonianTunnel}
\eeq
In the absence of  magneto-elastic coupling, the Hamiltonian 
\refe{eq:H_tot} was introduced in Refs. \onlinecite{glazman:1988,wingreen:1989}
to describe transport through a quantum dot in presence of a local electron-phonon 
coupling and later studied by several 
authors.\cite{koch:2005,koch:2006,braig:2003,mitra:2004,zazunov:2007,muhlbacher:2008}

\subsection{Combined coupling constant}
\label{combined}

The expression for the Hamiltonian can be further manipulated in such a way that 
the dimensionless coupling constants  $\phi_n$ and $\lambda_n$ introduced in subsections
\ref{magneto} and \ref{sec:polaron}, respectively, can be treated on the same 
footing.

For this purpose it is convenient to eliminate the linear polaron coupling in the
Hamiltonian \refe{eq:Hdot} by using the Lang-Firsov unitary transformation 
$H\rightarrow  UH U^{\dagger}$:
\begin{equation}
  \,  U=
e^{- \sum_n \lambda_n \left( b_n - b_n^{\dagger} \right) 
(n_{d\uparrow}+n_{d\downarrow})}
\,  .
\end{equation}
The transformation shifts the bosonic fields
($U b_n U^{\dagger} =  b_n - \lambda_n n_{d}$) giving for 
$ H_\mathrm{string} + H_{d}$:
\beq
{H}_{d} + {H}_{\mathrm{string}} \rightarrow  
\sum_n \omega_n^{\vphantom{\dagger}}  b_n^{\dagger}  b_n^{\vphantom{\dagger}} 
+ \sum_{\sigma} \tilde  \varepsilon_{d\sigma} n_{d\sigma} 
+ \tilde  U_{\infty} n_{d\uparrow} n_{d\downarrow} \, ,
\eeq
with $\tilde \varepsilon_{d\sigma}=\varepsilon_{d\sigma} - \sum_n \lambda_n^2  \omega_n$
and $\tilde U_\infty= U_{\infty} - 2  \sum_n \lambda_n^2  \omega_n$.
Both the energy of the dot level and  the
effective interaction between two electrons on the dot
are renormalized.
Note that, for  $U_{\infty} < 2 \sum_n \lambda_n^2 \omega_n$,
the effective interaction can become attractive.\cite{alexandrov:2003,koch:2005}
Focusing our analysis to strong Coulomb repulsion regime
we are not concerned by this possibility.

Using $U d_{\sigma} U^{\dagger} =d_{\sigma} e^{\sum_n \lambda_n (b_n-b_n^{\dagger})}$ 
and Baker-Hausdorff formula, the tunneling 
term in \refE{eq:HamiltonianTunnel} becomes
%s
\begin{eqnarray}
\label{eq:product}
& & U 
\left( e^{i\sum_n(b_n+b^\dagger_n)\phi_{\nu n}} d_\sigma \right) 
U^{\dagger} = e^{-i\sum_n \lambda_n \phi_{\nu n} } \qquad \qquad  \nonumber \\
& & \times 
e^{-2 i\sum_n \lambda_n \phi_{\nu n} n_{d\bar\sigma} }
\exp[\sum_n (\alpha_{\nu n} b_n -  \alpha_{\nu n}^* b_n^{\dagger} )]
d_\sigma \, ,
\end{eqnarray}
where $\bar \sigma$ indicate the spin projection opposite to $\sigma$.
In \refE{eq:product} we have introduced the {\it complex} 
electron-phonon coupling constants:
\begin{equation}
\alpha_{\nu n} =  \lambda_n + i \phi_{\nu n} \, .
\label{EquationTrenteSept}
\end{equation}

Let us first discuss the effect of the prefactor 
$\exp(- 2 i \sum_n \phi_{\nu n} \lambda_n n_{d\bar \sigma})$.
Since we are considering the limit where double occupied 
states are not accessible, it is easy to show that this term 
gives always 1 when evaluated with $d_\sigma$ on the 
three dot's basis states $\{|0\rangle, |\uparrow \rangle,|\downarrow \rangle \}$.
We will thus drop it in the following. 
The phase factor $ \exp(- \sum_n i \phi_{\nu n} \lambda_n) $ can 
be included in the definition of the tunneling amplitude 
$t_{\nu}$.

Combining the above results, we finally obtain the Hamiltonian
\begin{eqnarray}
H &=& \sum_n \omega_n  b_n^{\dagger}  b_n  
+ \sum_{\nu,k,\sigma} \xi_{\nu k} c^\dagger_{\nu k\sigma} c_{\nu k\sigma} 
+ \sum_{\sigma} \tilde{\varepsilon}_{d\sigma}  n_{d\sigma}  \nonumber \\
&+& \tilde U_{\infty}  n_{d\uparrow} n_{d\downarrow} + \tilde{H}_{T} \, . 
\label{eq:H_tot_1} 
\end{eqnarray}
with the tunneling Hamiltonian:
\begin{equation}
\label{eq:Htun_L}
\tilde{H}_{T}= 
\sum_{\nu, k,\sigma}
\{
t_{\nu}
\exp[\sum_n( \alpha_{\nu n} b_n - \alpha_{\nu n}^* b^\dagger_n)] 
c^\dagger_{\nu k\sigma} d_\sigma
+\mathrm{h.c.} 
\}.
\end{equation}

In \refE{eq:Htun_L} the polaronic and magneto-elastic couplings 
enter on the same footing through constants $\alpha_{\nu n}$,
see \refE{EquationTrenteSept}.

\subsection{Discussion}
\label{discussion}

%\subsubsection{Estimates of the coupling constants}

Remarkably, the ratio between the magneto-elastic and polaronic coupling constants,
cf. Eqs. \refe{eq:flux2} and \refe{eq:polaronic_coupling},
is independent of the length of the nanotube. A crude estimate shows that,
for the fundamental bending mode,
$\phi_0/\lambda_0\sim \sqrt{m/m^*}(e/C_g|V_g|) B h a_B/\Phi_0$, where
$a_B$ is the Bohr radius and
$m^*$ is the nucleon mass. (To get it we used $\rho\sim m^*/a_B$, $I\sim R_\perp^4\sim a_B^4$,
and $E\sim Ry/a_B^4$, where $Ry$ is the Rydberg energy.)
In the ratio $\phi_0/\lambda_0$, the smallness of $m/m^*$ is compensated by the largeness of $h/a_B$.
Thus, magneto-elastic and polaronic coupling constants of the same order of magnitude could be realized,
while being tunable independently through the magnetic field and gate voltage.
It is thus interesting to investigate their interplay in the 
transport properties of the device.

Let us now estimate the coupling constants quantitatively.
Assuming a linear mass density $\rho$ of the order of $10^{-15}$ kg/m,
corresponding to six carbon atoms/\AA,
and a length $L$ of 1 $\mu$m, we find that the typical mass of a 
single-wall carbon nanotube is $10^{-21}$ kg.
The zero point motion $X_{n0}$ for a mode 
of frequency $\omega_n/2\pi = 500$ MHz is $4$ pm.
For $B=40T$, \refE{eq:flux2} yields 
$\phi_0\approx 0.1$ for the fundamental bending mode 
[for which $\int_0^L (ds/L) f_0(s)\simeq 0.83$],
while, for $h=200$nm and $C_g V_g=5 e$, 
one finds from \refE{eq:polaronic_coupling} that $\lambda_0 \approx 0.1$.
The magnitude of the magneto-elastic and polaronic effects is
thus the same for these parameters.

%\subsubsection{Comments}

It is probably useful to point out that the current dependence generated 
by the strong magnetic field discussed in this paper and in previous 
works\cite{shekhter:2006,sonne:2009,rastelli:2010} is totally unrelated to 
the Aharonov-Bohm effect discussed in the literature in several
papers (see Ref. \onlinecite{charlier:2007} for a review).
The effect discussed there is due to the interference of the 
electrons in the Aharonov-Bohm loops generated by the carbon 
crystallographic structure.
The presence of phonon modes is not required for this effect to
be present, while it is crucial for the magnetoresistance 
discussed here.
It should be also noted that the two contributions are of course 
present at the same time in a realistic description of the 
carbon nanotube, but the two effects can be distinguished experimentally
since they have totally different thermal and voltage dependences.\cite{rastelli:2010}

Concerning the model for the description of the nanotube deformation,
in this paper we assumed that the standard theory of elastic 
continuous media applies.
This should be correct, at least for not too short nanotubes,
where an atomic description becomes necessary.
We only considered the case of a rigid rod with doubly clamped
boundary conditions. 
It is well know that it is possible to observe a crossover from this 
regime to the regime of tense string.\cite{sazanova:2004}
The effect of a residual stress on the string can be taken into account 
by modifying the Hamiltonian \refe{eq:H-string}.\cite{landau7}
The structure \refE{eq:Hamiltonian6} remains valid  in this case.
The effect of the tension modifies only the profile function $f_n(s)$ 
entering \refE{eq:flux} for the eigenmodes.
The elongation of the nanotube induced by the external gate and
magnetic field, with typical amplitude $\sim \lambda_n X_{n0}$ and
$ \phi_nX_{n0}$ respectively (we took $I\sim e \omega_n$ for the last estimate),
also generates an internal stress.
The nonlinear effects induced by this stress are safely neglected when
these amplitudes don't exceed $R_\perp$,\cite{landau7} which is the case for the parameters given above..

The amplitudes of the effective phases in the tunneling part of 
\refE{eq:Hamiltonian6} only coincide for symmetric boundary conditions. 
For instance, 
in the case of a nanotube that is clamped on one side and 
{\em hanging} on the other side, no specific relation would
exist between the phases in the left and right terms.

Surprisingly the Hamiltonian \refe{eq:H_tot_1}-\refe{eq:Htun_L}
with $\lambda_n=0$ 
has been already introduced before 
(in the spinless case) for a quite different problem.
Imam {\em et al.}\cite{imam:1994} considered 
resonant tunneling in presence of 
an electromagnetic environment associated with 
the fluctuations of the bias voltage, 
and described by a bath of phonon modes.
Nevertheless, the effect of the environment on the conductance 
was only addressed at large bias voltage or far from the resonance in that work.
Moreover in this paper we address the interplay of the magnetic
and polaronic coupling, that was not present there.

%%%%%%%%%%%%%%%%%%%%%%%%%%%%%
%
%			Section III FORMALISM
%
%%%%%%%%%%%%%%%%%%%%%%%%%%%%%

\section{formalism}
\label{formalism}

Our aim is to derive the current-voltage characteristics for the problem described 
with Hamiltonian \refe{eq:H_tot_1}-\refe{eq:Htun_L}.
(Note that tildes that appear there will be omitted from now on.)
Here we present the general approach that will be used in the next Sections.

The problem at hand is a generalization of the well studied 
polaron problem.
It has not been solved exactly yet, to the best of our knowledge.
Nevertheless, it is possible to obtain approximate results in different regimes.
When the coupling constant is small, a perturbative expansion is possible;
it has been performed both for the polaronic\cite{egger:2008} and 
the magneto-elastic cases.\cite{rastelli:2010}
In principle the method used in those previous publications can 
be extended to investigate the case when both couplings are 
present.
However it works only when the predicted effects are small
and we will not use it in this paper.
Alternatively, the dynamics of a single
electron in the presence of the phonon bath can be 
obtained.\cite{glazman:1988,imam:1994} 
However, the results obtained for the phonon-assisted resonant tunneling can 
only be applied far from the resonance, when effects associated with the Fermi
sea in the leads can be neglected.
Another tractable limit is the case of high-temperature incoherent tunneling
limit ($T\gg \Gamma$),
which has been discussed in Ref. \onlinecite{koch:2006} for the polaron
problem and does not require that the electron-phonon coupling is perturbative.
Here, we generalize this approach to describe the 
effect of the interplay of the two couplings.

In this incoherent regime the electronic environment at temperature $T$ 
has the time to suppress the coherence of the quantum evolution 
between tunneling events. 
The equations of motion for the off-diagonal elements of the density matrix 
in the basis of the eigenvectors of $H_0=H-H_T$ decouple from those 
for the diagonal elements. 
It is then possible to obtain the equations of motion for the diagonal elements
({\em i.e.} the occupation probabilities) alone.

The rates can be obtained by standard perturbation theory.
The time evolution of any eigenstate $|i\rangle$ of $H_0$
can be written in terms of the resolvent:
\beq
	|i\rangle(t) = \int {dE \over 2 \pi} {e^{-i E t} \over E -H_0 -H_T+i\eta} 
	|i\rangle
	\,.
\eeq
The probability that the system is in another state $|f\rangle$ after a time $t$ is 
\beq
	P_{if} 
	= 
	\left|
	\int {dE \over 2 \pi} 
	e^{-i E t}
	\langle f| G^0(E) + G^0(E) T(E) G^0(E)  |i\rangle
	\right|^2
	\,,
	\label{Pif}
\eeq
where $T(E)=H_T+G^0(E)H_TG^0(E)+\dots$ is the $T$-matrix 
and $G^0(E)=(E-H_0+i\eta)^{-1}$ is the free retarded propagator.
Neglecting the $E$ dependence of $T(E)$ in \refE{Pif} gives the 
standard Fermi golden rule with $T(E_i)$ replacing $H_T$, that is,
$P_{if} = t W_{if}$ with 
\beq
	W_{if} = 2\pi |T_{if}(E_i)|^2\delta(E_i-E_f)
	\label{golden}
	\,,
\eeq
and $E_i$ the eigenvalue of $H_0$ related to the state $i$.
By inserting the resolution of the identity in terms of 
the eigenstates of $H_0$, one can then write for the first two orders:
\beq
	|T_{if}(E_i)|^2 = \left|(H_T)_{if}  + \sum_n { (H_T)_{in} (H_T)_{nf} \over E_i-E_n}  
	\right|^2
	\,.
	\label{Tmat}
\eeq
These two terms give the tunneling and cotunneling contributions to the transition rate,
respectively.
We will investigate the effect of the tunneling term
in Section \ref{sequential} and then the corrections given by the cotunneling terms 
in Section \ref{cotunneling}.
A comment is at order. The expression \refe{golden} holds if the frequency dependence 
of $T(\omega)$ is sufficiently smooth. This is not true if one intermediate 
state has the same energy as the final state. 
The problem has been discussed in Ref. \onlinecite{koch:2006} and it requires 
a proper renormalization of the divergences of the integrals appearing 
in the calculation of the transition rates.
The regularization for the transition
rates associated with cotunneling processes is done in Appendix \ref{app:W_nm}.

The system state is thus completely described by the probability 
of being in one of the eigenstate of $H_0$. 
We will assume that the lead remains in thermal equilibrium
during the evolution, thus the only relevant degrees of freedom
left are the dot and oscillator degree of freedom. 
In the following we will focus on a single phononic mode, 
for which we define $\omega$ the resonance frequency,
$\phi$ the magneto-elastic coupling, and $\lambda$ 
the polaronic coupling.
In order to fully describe the system we finally define the 
probability $p_{sn}(t)$ that the dot is in one of the three available 
states $\left(| s\rangle = \left| 0 \right> , \left| \downarrow  \right>,
\left| \uparrow \right>  \right)$ and the oscillator in the phonon
state $n$ ($| n\rangle={b^\dag}^n |0\rangle/\sqrt{n!}$). 
The rate equations read:
\begin{eqnarray}
\frac{d p_{sn}}{d t} &=& \sum^{\infty}_{m=0} \sum_{s'=0,\uparrow,\downarrow}
\left[ W_{s'm,sn} \; p_{s'm}
-   W_{sn,s'm} \;  p_{sn} \right] \nonumber \\
&-& \frac{1}{\tau} \left( p_{sn} - P^{(\mathrm{eq})}_{n}\sum^{\infty}_{m=0} p_{sm} \right) \, .
\label{eq:rateseqs}
\end{eqnarray}
Here $W_{sn,s'm}$ is  the transition rate  
from the electronic dot state $s$ and phonon state $n$
to another configuration $(s',m)$.
Following Ref.~\onlinecite{koch:2006}, a phenomenological 
relaxation term has been introduced in \refE{eq:rateseqs} 
to describe the coupling of the oscillator 
with the non-electronic environment at the origin of an intrinsic finite 
quality factor $Q^{-1}=\omega \tau/2\pi $ for the oscillator.
The function $P^{(\mathrm{eq})}_n$ gives the equilibrium phonon distribution function:
\beq
	P^{(\mathrm{eq})}_n =  (1-e^{-\omega/ T}) e^{-n\omega / T} \, .
\eeq

Given the rates, equations \refe{eq:rateseqs} can be solved; the average
current is then obtained from the stationary solution for the probabilities.
In the next two sections, we discuss the result for the current 
at the sequential (Sec. \ref{sequential}) and cotunneling (Sec. \ref{cotunneling}) orders.

%%%%%%%%%%%%%%%%%%%%%%%%%%%%%
%
%			Section IV  SEQUENTIAL TUNNELLING REGIME
%
%%%%%%%%%%%%%%%%%%%%%%%%%%%%%

\section{Sequential Tunneling Regime}
\label{sequential}

The sequential tunneling regime corresponds to keeping only the 
 lowest order term in the expansion \refe{Tmat} that determines
the transition rates.
In this regime we find that the current-voltage characteristics depends on
a single coupling constant which does not allow to distinguish the 
polaronic and magneto-elastic effect from each other.

\subsection{General formula}
The explicit form of the transition rates in the sequential 
tunneling limit reads:
\beq
	W_{sn,s'm} = \sum_{\nu=l,r} 
	W^{(\nu)}_{sn,s'm}
\eeq
with
\beqa
	W^{(\nu)}_{sn,s'm}
	&=&
	{2\pi} 
		\sum_{i_{\nu},f_{\nu}} P^{(\mathrm{eq})}(i_{\nu}) 
	{\left| \langle f_{\nu},s',m \right| {H}_T \left| i_{\nu},s,n \rangle \right|}^2
\nonumber \\
	&&\quad \times \delta( E_{f_{\nu}} -  E_{i_{\nu}} + \varepsilon_{ds'}-\varepsilon_{ds} + \omega [m-n]) 
	\,.
\label{eq:Wseqtunn}
\nonumber \\
\eeqa 
Here $i_\nu$ and $f_\nu$ are initial and final fermionic states in  lead $\nu$,
with energies $E_{i_{\nu}}$ and $E_{f_\nu}$, 
$P^{(\mathrm{eq})}(i_{\nu})$ is  the equilibrium distribution of state $i_\nu$, and
for convenience we introduced $\varepsilon_{d0}\equiv 0$.
The non vanishing rates take the form:
\begin{eqnarray}
W^{(\nu)}_{0n,\sigma m} &=& \!\!
 \Gamma_{\nu} {\left| M_{nm}^{(\nu)} \right|}^{2}
n_{\nu} \left(  \varepsilon_{d\sigma}  +  \omega [m-n]   \right) ,\label{eq:Wseq1} \\
W^{(\nu)}_{\sigma m,0n}  &=& \!\! 
 \Gamma_{\nu}  {\left| M_{nm}^{(\nu)} \right|}^{2} 
 \left[
1 - n_{\nu} \left( \varepsilon_{d\sigma}  + \omega [m-n]   \right)  
\right]  \, ,
\nonumber \\ \label{eq:Wseq2}  
\end{eqnarray}
where
$n_{\nu}(\varepsilon)=1/[e^{(\varepsilon-\mu_\nu)/ T}+1]^{-1}$ 
is the Fermi distribution in lead $\nu$, 
$M^{(\nu)}_{nm}=\langle n| e^{\alpha_\nu b-\alpha_\nu^*b^\dag)}|m\rangle$,
and we recall that $\Gamma_\nu=2\pi |t_\nu|^2 N_\nu$.

The absolute value of the matrix element $M^{(\nu)}_{nm}$
depends only on $|\alpha_\nu|\equiv \alpha$, where
\beq
\alpha=\sqrt{\lambda^2+\phi^2}
\eeq
is the r.m.s. of the polaronic and magneto-elastic coupling
constants.
This can be readily be shown by absorbing  the 
phase of $\alpha_\nu$  in the operators $b$ and $b^\dag$
appearing in the definition of $M^{(\nu)}_{nm}$. 
This is a surprising results, which holds only at this sequential order in the tunneling expansion.
For instance, it was shown in Ref. \onlinecite{rastelli:2010} that there are qualitative differences 
between the magneto-elastic and the polaronic problems:
at resonance, the maximum current depends on the asymmetry of the
electronic coupling between the dot and the leads
in the first case and does not in the second case.
It is thus clear that to higher order differences  
have to appear. 
Technically this means that the current should depend
also on the phase of the coupling $\alpha_\nu$,
as we will find at the cotunneling level in Sec. \ref{cotunneling}

In the reminder of this Section we will study the evolution 
of the current-voltage characteristics as a function of 
the magnetic field. 
For simplicity we assume that the relaxation time introduced in
\refE{eq:rateseqs} is shorter than the typical tunneling rate, so that 
the distribution function can be written:
\begin{equation}
p_{sn}^{\vphantom{\dagger}} = P_s^{\vphantom{\dagger}}  
P^{(\mathrm{eq})}_n  
\label{eqn:P_eq_sn}
\,.
\end{equation}
The rate equation \refe{eq:rateseqs} for the stationary solution
($dp_{sn}/dt=0$) can then be cast in the form
${\bf W} {\bf P} = 0 $, 
where ${\bf P}=(P_0,P_{\uparrow},P_{\downarrow})$
(with the normalization condition $P_0+P_{\uparrow}+P_{\downarrow}=1$),
\begin{equation}
{\bf W} = \sum_{\nu=l,r}
\left(
\begin{array}{ccc}
-\left[W^{(\nu)}_{0\uparrow}+W^{(\nu)}_{0\downarrow}\right] &  W^{(\nu)}_{\uparrow0} &  W^{(\nu)}_{\downarrow0} \\
  W^{(\nu)}_{0\uparrow}                  & -W^{(\nu)}_{\uparrow0} & 0 \\
  W^{(\nu)}_{0\downarrow}                & 0 & -W^{(\nu)}_{\downarrow0}
\end{array}
\right) \, ,
\label{eq:rateseqs_3}
\end{equation}
and we define  the {\em reduced} transition rates: 
\beq
W^{(\nu)}_{ss'} =
\sum_{n,m} P^{(\mathrm{eq})}_n
W^{(\nu)}_{sn,s'm} \,.
\eeq
By performing the sum over the phonon states the rate take a form similar to the
non interacting case:\cite{braig:2003} 
$W^{(\nu)}_{0\sigma} =  \Gamma_{\nu}  \; \tilde{n}_{\nu}(\varepsilon_{d\sigma})$
and 
$W^{(\nu)}_{\sigma0} =  W^{(\nu)}_{0\sigma} e^{(\varepsilon_{d\sigma} - \mu_{\nu})/ T}  $
with
\beq
	\tilde{n}_{\nu}(\varepsilon) 
	= \int dE \; {\cal P}(E) \; n_{\nu}(\varepsilon+ E)
	\,.
	\label{equationntilde} 
\eeq
Here the function ${\cal P}$ reads
\beq
{\cal P}(E) = \frac{1}{2\pi}
\int dt
e^{i E  t}  {\left< A_\nu(t)A_\nu^{\dagger}(0) \right>}  \, ,  
\eeq
with
$
A_\nu(t) = e^{\alpha_{\nu} b(t) - \alpha_{\nu}^* b^{\dagger}(t)}
$.
In Fourier space, ${\cal P}(E)$ takes the form of a  
a sum of delta functions centered at integers multiples 
of the oscillator frequency $\omega$. 
In the limit $ T\ll \omega$, on which we will focus hereafter, 
the oscillator is in the ground state and 
\beq
\label{eq:PE}
   {\cal  P}(E) 
	= 
	\sum_{m=0}^{\infty} e^{-\alpha^2} {\alpha^{2m}\over m!}
	\delta(E-m\omega)
	\,.
\eeq

The current, say for instance at the left lead, 
can be written in terms of the
probabilities and the tunneling rates:
\begin{equation}
I = e \sum_{\sigma=\uparrow,\downarrow} \sum^{\infty}_{n,m=0}
\left(
W^{(l)}_{0n,\sigma m} p_{0n}
-
W^{(l)}_{\sigma m,0n} p_{\sigma m}
\right)  \, . \label{eqn:Current}
\end{equation}
In the case of interest here of the equilibrated oscillator
one obtains: 
\begin{equation}
I  =   \sum_{\sigma=\uparrow,\downarrow}
\left( 1 - P_{\bar\sigma} \right) I_{\sigma} \, , \label{eqn:Iseq_1} 
\end{equation}
with the current for each spin-channel: 
\begin{equation}
I_{\sigma}
=e \Gamma_l \Gamma_r
\frac{
\tilde{n}_{l}(\varepsilon_{d\sigma})  \tilde{n}_{r}(\varepsilon_{d\sigma})
\left[
e^{(\varepsilon_{d\sigma}-\mu_r)/T}
-
e^{(\varepsilon_{d\sigma}-\mu_l)/T}
\right]
}
{\sum_{\nu} \Gamma_{\nu} \tilde{n}_{\nu}(\varepsilon_{d\sigma})
/ n_{\nu}(\varepsilon_{d\sigma})} \, , \label{eqn:Iseq_2} 
\end{equation}
and the stationary probabilities:
\begin{equation}
P_{\sigma} =
\frac{\sum_{\nu,\nu'} \Gamma_{\nu} \Gamma_{\nu'}
\tilde{n}_{\nu}(\varepsilon_{d\sigma}) \tilde{n}_{\nu'}(\varepsilon_{d\bar\sigma})
e^{(\varepsilon_{d\bar\sigma}-\mu_{\nu'})/T}}
{\sum_{\nu,\nu'} \Gamma_{\nu} \Gamma_{\nu'}
\tilde{n}_{\nu}(\varepsilon_{d\uparrow}) \tilde{n}_{\nu'}(\varepsilon_{d\downarrow})
\left[
 n^{-1}_{\nu}(\varepsilon_{d\uparrow}) n^{-1}_{\nu'}(\varepsilon_{d\downarrow}) - 1
\right]} \, , \label{eqn:Ps_s}
\end{equation}
with $P_0=1-P_\uparrow-P_\downarrow$.
Formula \refe{eqn:Iseq_1} for the sequential current has a simple interpretation:
the current associated to the tunneling through
one spin state and in absence of any correlation between the electrons
is reduced by a factor which is just the probability of occupancy
of the dot with opposite spin. 
Indeed, $I_{\sigma}$ corresponds exactly to the sequential current 
for the spinless problem in presence of electron-phonon interaction.

From the previous expressions, we can obtain the  
linear conductance in the sequential tunneling regime:
\begin{equation}
G = G_0
\frac{\Gamma}{T}
%\left(
\frac{
\tilde{n}(\varepsilon_{d\uparrow})
+
\tilde{n}(\varepsilon_{d\downarrow})
}{1 +
2 e^{-(\varepsilon_d-\mu)/{T}} \cosh({\mu_BB}/{T})}
%\right) 
\, , \label{eqn:Gs}
\end{equation}
where $G_0= e^2 \Gamma_l\Gamma_r / \Gamma^2$,
$\eps_d=(\eps_{d\uparrow}+\eps_{d\downarrow})/2$ and
$\mu=(\mu_{l}+\mu_{r})/2$.  
Eqs. \refe{eqn:Iseq_1}-\refe{eqn:Gs} are the central result of this Section.
In the following we discuss the current
in different regimes.

\subsection{Linear conductance}

The linear conductance generically displays a resonance as the energy
$\eps_d$ of the discrete level in the dot is varied, e.g. with a gate voltage. 

In the absence of electron-phonon interaction,\cite{glazman:1988b} the conductance is peaked
around $\eps_d-\mu\simeq(T/2)\log 2$, with width $\sim T$ and maximum height
$G_\mathrm{max}=2G_0(\Gamma/T)/(3+2\sqrt{2})$ at zero magnetic field.
It is peaked around around $\eps_d-\mu\simeq \mu_B B$ (assuming $B>0$), with width $\sim T$ and maximum height
$G_\mathrm{max}=G_0\Gamma/(4T)$ at large magnetic field $\mu_B B\gg T$.

The main effect of electron-phonon interactions is to suppress the peak height: 
At low temperature ($T\ll\omega$), we find with
help of Eqs. \refe{equationntilde}, \refe{eq:PE} that the suppression factor is $e^{-\alpha^2}$. \cite{glazman:1988,braig:2003}

\subsection{Non-linear $IV$ characteristics}

In this section, we assume that the bias voltage is applied symmetrically, 
with $\mu_{l,r}=\mu\pm eV/2$.

\begin{figure}[thbp]
\includegraphics[scale=0.35,angle=270.]{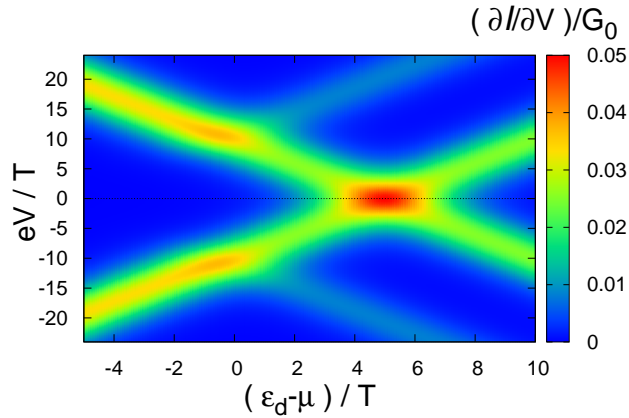}
\caption{Differential conductance in the absence of electron-phonon coupling
($\alpha=0$)
%scaled with $G_0$ 
as a function of gate and bias voltages in the sequential tunneling regime, 
for $\mu_B B/T  = 5$, $\Gamma_l/T=\Gamma_r/T=0.1$.}
\label{fig2_alpha=0} 
\end{figure}

In the absence of electron-phonon interaction and at zero magnetic field,
resonance lines at $\eps_d-\mu\simeq\pm eV/2$, with width $\sim T$, 
separate the regions where the current is blocked and those where
it reaches a finite value $I=2e\Gamma_l\Gamma_r/(\Gamma+\Gamma_l)$
or $I=-2e\Gamma_l\Gamma_r/(\Gamma+\Gamma_r)$, depending
on the sign of the bias voltage.\cite{glazman:1988b}
At finite magnetic field, the regions where the current is blocked and those
where down spins only can flow through the dot are
separated by the lines $\eps_d-\mu-\mu_BB\simeq\pm eV/2$.
Inside the conducting regions it appear bands where an additional conduction channel
opens for up spins.
At $V>0$ (respectively $V<0$), these bands stand between the lines
$\eps_d-\mu -eV/2=\pm \mu_BB$ ($\eps_d-\mu +eV/2=\pm \mu_BB$);
the current reaches $I=e\Gamma_l\Gamma_r/\Gamma$
($I=-e\Gamma_l\Gamma_r/\Gamma$) inside the bands.
The contrast between the different lines in the gate and bias voltage dependence
of the differential conductance, see Fig. \ref{fig2_alpha=0}, 
is readily explained by the amplitudes for the current
found above between the lines.
%
%%%%%%%%%%%%%%%%%%%%%%%%%%%%%%
%
%				Figure 2
%
%%%%%%%%%%%%%%%%%%%%%%%%%%%%%%
%
%%%%%%%%%%%%%%%%%%%%%%%%%%%%%%%

\begin{figure}[bhtp]
\includegraphics[scale=0.35,angle=270.]{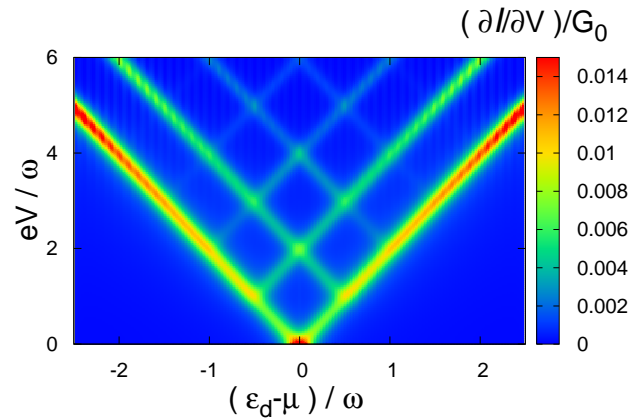}
\caption{Differential conductance at zero magnetic field as a function of gate and bias voltages in
the sequential tunneling regime, for $\alpha=1$, %$B=0$, 
$\omega/T=20$, and
$\Gamma_l/T=\Gamma_r/T=0.1$.}
\label{fig3_B=0}
\end{figure}

The case of finite electron-phonon coupling and  zero magnetic field was discussed 
in details in Refs.\onlinecite{braig:2003},\onlinecite{koch:2005}. 
Note that Equation \refe{eqn:Iseq_2}  for the current simplifies into
\begin{equation}
I  = 2e\Gamma_l\Gamma_r\frac{  
\tilde{n}_{l}(\varepsilon_{d})  \tilde{n}_{l}(\varepsilon_{d})
\left(
e^{(\varepsilon_{d}-\mu_R)/k_B T}
-
e^{(\varepsilon_{d}-\mu_L)/k_B T}
\right)}
{\sum_{\nu} \Gamma_{\nu}
\tilde{n}_{\nu}(\varepsilon_{d})
\left(n^{-1}_{\nu}(\varepsilon_{d})+1 \right)} \, .
\end{equation}
The main features of the current are (i)  the  presence of inelastic lines
(vibrational sidebands)
at $\eps_d-\mu +n\omega=\pm eV/2$ in the conducting regions and 
(ii) a suppression of the current at the low bias voltage
(Frank-Condon blockade) when the electron-phonon coupling is large enough. 
Fig. \ref{fig3_B=0} illustrates point (i)  for intermediate 
electron-phonon coupling.
%
%%%%%%%%%%%%%%%%%%%%%%%%%%%%%%
%
%				Figure 3
%
%%%%%%%%%%%%%%%%%%%%%%%%%%%%%%
%
%%%%%%%%%%%%%%%%%%%%%%%%%%%%%%%

\begin{figure}[htbp]
\includegraphics[scale=0.35,angle=270.]{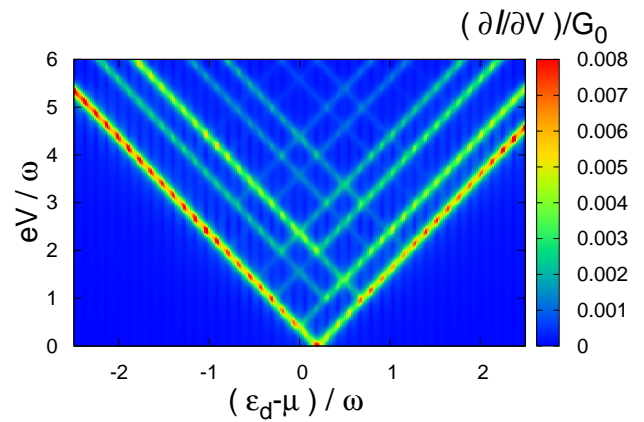}
\caption{Differential conductance as a function of gate and bias voltages in
the sequential tunneling regime, for  $\lambda=1$, $\phi=0.4$, $\mu_BB/\omega=0.2$, 
$\omega/T=40$, and $\Gamma_l/T=\Gamma_r/T=0.1$.}
\label{fig4_Is_case1}
\end{figure}

\begin{figure}[h!]
\includegraphics[scale=0.32,angle=270.]{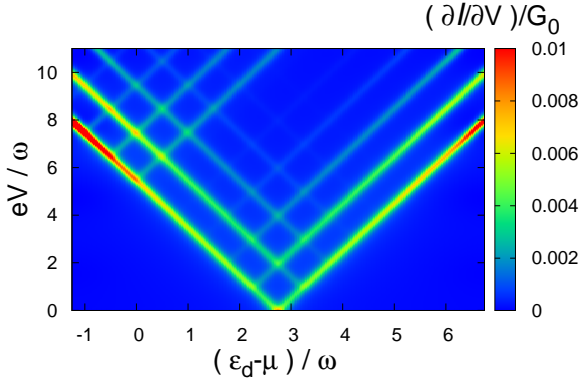}
\caption{Differential conductance as a function of gate and bias voltages in
the sequential tunneling regime, for  $\lambda=1$, $\phi=0.4$, $\mu_BB/\omega=2.75$, 
$\omega/T=20$, and $\Gamma_l/T=\Gamma_r/T=0.1$.}
\label{fig5_Is_case2}
\end{figure}

We can now understand the general features of the current when both 
the electron-phonon coupling and magnetic field are present.
They look different depending on the ratio $\mu_B B/\omega$.
When $\omega>\mu_BB$, the vibrational sidebands are split due to the removal of the spin
degeneracy of the single level, see Fig. \ref{fig4_Is_case1}.
When $\mu_BB>\omega$, vibrational sidebands are now present as replicas of the 
two main resonance lines, see Fig. \ref{fig5_Is_case2}.
%
%
%
%%%%%%%%%%%%%%%%%%%%%%%%%%%%%%
%
%				Figure 4
%
%%%%%%%%%%%%%%%%%%%%%%%%%%%%%%
%
%%%%%%%%%%%%%%%%%%%%%%%%%%%%%%%
%
%%%%%%%%%%%%%%%%%%%%%%%%%%%%%%
%
%				Figure 5
%
%%%%%%%%%%%%%%%%%%%%%%%%%%%%%%
%
%%%%%%%%%%%%%%%%%%%%%%%%%%%%%%% 

In conclusion of this Section, we found that the polaronic and magneto-elastic couplings
cannot be distinguished in the sequential tunneling regime,
apart from the removal of the spin degeneracy
induced by the Zeeman effect. 
In the next Section, we will obtain a different
conclusion when higher order processes in the tunneling expansion 
are considered.

%%%%%%%%%%%%%%%%%%%%%%%%%%%%%
%
%			Section V  COTUNNELLING
%
%%%%%%%%%%%%%%%%%%%%%%%%%%%%%

\section{Cotunneling}
\label{cotunneling}

We first discuss the general approach to consider the cotunneling corrections 
to the  rate equations.
Then we show that, in a specific regime,  
the cotunneling process dominates the total current.
Finally, we concentrate on the elastic 
tunnel current to clarify the differences 
between the  magneto-elastic and polaronic interactions. 

\subsection{General formula}

The explicit form of the transition rates induced by cotunneling processes 
is given in Appendix \ref{app:W_nm}.

Assuming that the phonons are equilibrated, we find that
the rate equation \refe{eq:rateseqs} including cotunneling events reads
\beq
\label{eq:rateseqs_10}
({\bf W} +\delta {\bf W} ) ({\bf P}+\delta {\bf P}) = 0,
\eeq
where $\delta{\bf P}$ is a correction to the occupation probabilities  ${\bf P}$
found in Sec. \ref{sequential} 
(with $\delta P_0+\delta P _\uparrow+ \delta P_\downarrow=0$)
and
\begin{equation}
\delta {\bf W} = \sum_{\nu,\mu=l,r}
\left(
\begin{array}{ccc}
 0  &  0 &  0 \\
 0  & -W^{(\mu \nu)}_{\uparrow\downarrow} &  W^{(\mu \nu)}_{\downarrow\uparrow} \\
 0   & W^{(\mu \nu)}_{\uparrow\downarrow} & -W^{(\mu \nu)}_{\downarrow\uparrow}
\end{array}
\right)
\label{eq:rateseqs_4}
\end{equation}
and
we introduced the reduced cotunneling rates
\beq
W^{(\nu\nu')}_{ss'}
=
\sum_{n,m} P^{(\mathrm{eq})}_n
W^{(\nu \nu')}_{sn,s'm}  \, .
\eeq
corresponding to the transmission of one electron from  lead $\nu$ to lead
$\nu'$ without variation of the dot state, $s=s'=0,\uparrow,\downarrow$, or with
a spin-flip process between $\sigma=\uparrow$ and $\sigma'=\downarrow$ 

Considering that the cotunneling rates are smaller than
the sequential tunneling rates by a factor $\sim \Gamma /T$,
we neglect a term $\delta {\bf W} \delta {\bf P}$ in \refE{eq:rateseqs_10}
and  we find that the correction to the occupation probabilities solves
${\bf W} \delta {\bf P}=-\delta {\bf W} {\bf P}$.
Note that, when the oscillator is at thermal equilibrium, the cotunneling
processes affect the steady occupation probabilities ${\bf P}$ only via the
spin-flip processes.
In the limit of vanishing magnetic field, we have, by symmetry, 
$W^{(\mu \nu)}_{\uparrow\downarrow}=W^{(\mu \nu)}_{\downarrow\uparrow}$
and $P_{\uparrow}=P_{\downarrow}$.
Then the cotunneling rates  completely cancel from \refE{eq:rateseqs_10},
in agreement with the results of Ref. \onlinecite{koch:2005}.

According to the previous discussion, 
the cotunneling correction to the current can be decomposed in two parts: 
\begin{equation}
\label{eqn:Itot}
\delta I= \delta I_{1} + \delta I_{2} \, .
\end{equation}
They represent respectively  the correction to the current
due to the electron transfer via cotunneling:
\begin{eqnarray}
\delta I_{1} &=& e \left(
W^{(lr)}_{0 0}
-
W^{(rl)}_{0 0}
\right)   P_{0}  \nonumber \\
&+&
e \sum_{\sigma,\sigma'=\uparrow,\downarrow}
\left(
W^{(lr)}_{\sigma \sigma'}
-
W^{(rl)}_{\sigma \sigma'}
\right)  P_{\sigma} \label{eqn:Icot_1} \, , 
\end{eqnarray}
and the effect of the variation of the occupation probabilities produced by the
cotunneling  processes:
\begin{equation}
\label{eqn:Icot_2}
\delta I_{2}  = e \sum_{\sigma} \left(
W^{(l)}_{0 \sigma} \delta P_{0}
-
W^{(l)}_{\sigma 0} \delta P_{\sigma}
\right)
\end{equation}
In the next Section, we discuss the results for the current.

\subsection{Discussion}

\begin{figure}[b]
\includegraphics[scale=0.3,angle=270.]{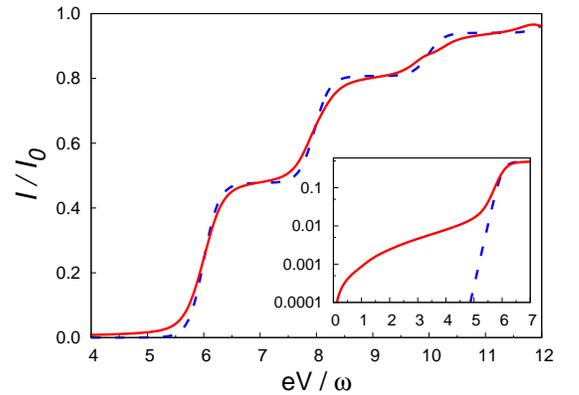}
\caption{Bias voltage dependence of the current with (straight line) 
and without (dashed line) the cotunneling contribution, 
for $\lambda=1$, $\phi=0.4$,
$ \Gamma_l/T = \Gamma_r/T = 0.4$,
$\omega/T = 15 $, 
$(\varepsilon_d-\mu)/ \omega=5$, and
$\mu_B B/  \omega=2$. Inset: 
Zoom of the low-voltage region in log-scale.
}
\label{fig6}
\end{figure}

The above formalism allows us to obtain the current including both sequential 
tunneling and cotunneling contributions, for arbitrary parameters of the junction.
An example is shown in Fig. \ref{fig6}.
In general, cotunneling processes produce a smoothing of the features already
discussed in the sequential current, that is the steps associated with the vibrational sidebands and the Zeeman splitting.
%
%%%%%%%%%%%%%%%%%%%%%%%%%%%%%%
%
%				Figure 6
%
%%%%%%%%%%%%%%%%%%%%%%%%%%%%%%
%
%%%%%%%%%%%%%%%%%%%%%%%%%%%%%%%
%
However they dominate the current in the regions where it is strongly suppressed at the sequential level, see inset of  Fig. \ref{fig6}.

In the following, we will concentrate on the region where the cotunneling terms dominate the current.
To simplify  the expressions for the current, we will further assume that the gate voltage is tuned in the region
where the dot is mostly empty ($P_0\approx 0$) and the magnetic field is sufficiently large that only
the down spin state for the singly occupied dot can be accessed virtually,
that is when $\eps_d-\mu,\mu_B B\gg T,eV$. In addition, we will consider that $T\ll\omega$ so that the
oscillator is in its ground state.
Then the expression \refe{eqn:Itot} for the current 
(noted $I$ instead of $\delta I$ from now)
simplifies to
\begin{equation}
\label{eqn:I_tun_P0}
I = e \left( W^{(lr)}_{00} - W^{(rl)}_{00} \right) \, ,
\end{equation}
with
\begin{eqnarray}
W^{(\nu\nu')}_{00} & =&
\frac{\Gamma_{\nu}\Gamma_{\nu'}}{2\pi}
\sum^{\infty}_{m,p,q=0}
M^{(\nu)*}_{0q}  M^{(\nu')}_{mq}  M^{(\nu)}_{0p}  M^{(\nu')*}_{mp}
\nonumber\\
&&\times\mathcal{F}
\left( \mu_{\nu} ; \mu_{\nu'}  +m\omega ;
\varepsilon_{d \downarrow} +q\omega;
\varepsilon_{d \downarrow}+p\omega
\right).
\nonumber\\
\label{eqn:I_tun_0_d}
\end{eqnarray}
where the function $\mathcal{F}$ is defined in Appendix B.
A finite electron-phonon interaction results in the suppression of the cotunneling current,
compared to the current in the absence of that interaction
$I^{(0)}= (I_0\Gamma/4 \pi^2 T)
\{
\Psi'[1/2 +i ({\mu_l - \varepsilon_{d\downarrow}})/({2\pi T})] 
-\Psi'[1/2 +i ({\mu_r - \varepsilon_{d\downarrow}})/({2\pi T})] ) 
\}
$,
where $I_0=e\Gamma_l\Gamma_r/\Gamma$.  
Similarly to the results found in Sec \ref{sequential}
we find that the polaronic and magneto-elastic couplings cannot be distinguished 
close to the charge degeneracy point, at $\eps_{d\downarrow}-\mu\ll\omega$.
Focussing for instance on the elastic contribution to the cotunneling current, 
\beq
I=I^{(0)}|M_{00}^{(l)}|^2|M_{00}^{(r)}|^2=I^{(0)}e^{-2\alpha^2},
\eeq
we find that the suppression factor is the square of the one found in the sequential tunneling regime.
This effect can be understood by (i) projecting Hamiltonian \refe{eq:H_tot_1}-\refe{eq:Htun_L} 
over the oscillator's ground state,\cite{rastelli:2010}
thus obtaining an effective Hamiltonian for a non-interacting resonant level with renormalized tunneling amplitudes
$t_\nu^{\mathrm{eff}}=t_\nu\exp{(-\alpha^2/2)}$, and (ii) recalling that the current is obtained as a fourth-order
effect in these tunneling amplitudes.

Instead, far away from the charge resonance at $\eps_{d\downarrow}-\mu\gg\omega$, 
we find 
\beqa
\label{eq:I_tun_el_ad}
I & =&  I^{(0)}
\sum^{\infty}_{q,p=0}
M^{(l)*}_{0q}  M^{(r)}_{0q}  M^{(l)}_{0p}  M^{(r)*}_{0p}
\nonumber\\
&&=  I^{(0)} 
|\langle 0|A_lA_r^\dagger|0\rangle|^2
=I^{(0)}e^{-2\phi^2}
\,
\eeqa
(here $A_\nu=e^{\alpha_\nu b-\alpha^*_\nu b^\dag}$),
which is only suppressed by the magneto-elastic coupling.
The  absence of an exponential suppression of the cotunneling current in the polaronic case 
was noted in Ref. \onlinecite{koch:2006}. 
The magneto-elastic effect can be understood by accounting for the virtual transitions to
the singly occupied states of the dot perturbatively.\cite{shekhter:2006}

\begin{figure}[h]
\includegraphics[scale=0.3,angle=270.]{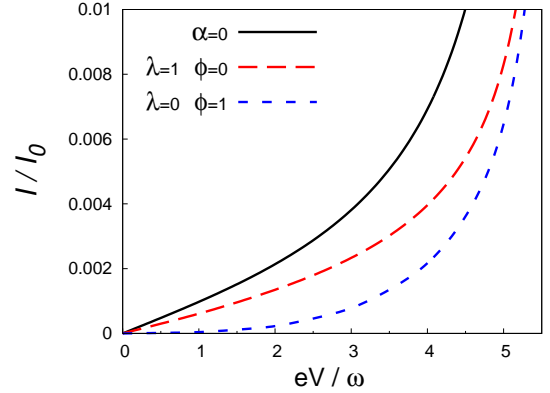}
\caption{Bias voltage dependence of the current for different electron-phonon couplings and 
$ \Gamma_l/T = \Gamma_r/T = 0.4$,
$\omega/T = 15 $, $(\varepsilon_d-\mu)/\omega=5$, and
$\mu_B B/\omega =2$.}
\label{fig18}
\end{figure}

In general, the polaronic and magneto-elastic couplings suppress
the cotunneling current in different ways at not too large oscillator's frequency.
The difference is traced back to the way the couplings appear in
the tunneling Hamiltonian \refe{eq:Htun_L}. 
In the polaronic case, an electron crossing the junction is subject to opposite phases
$\pm i\lambda_n(b_n-b^\dagger_n)$ at the tunnel barriers with the leads and the dot,
which can result in a compensation of their effect.
Instead, in the magneto-elastic case the electron feels the same phase $\phi(b+b^\dagger)$
at each barrier.
Indeed, for the same amplitude of the coupling constants, we find that the suppression 
of the cotunneling current is
stronger for the magneto-elastic interaction 
than for the polaronic interaction, see Fig. \ref{fig18}. 

%
%%%%%%%%%%%%%%%%%%%%%%%%%%%%%%
%
%				Figure 9
%
%%%%%%%%%%%%%%%%%%%%%%%%%%%%%%
%
%%%%%%%%%%%%%%%%%%%%%%%%%%%%%%% 

%%%%%%%%%%%%%%%%%%%%%%%%%%%%%
%
%			Section VI  CONCLUSIONS
%
%%%%%%%%%%%%%%%%%%%%%%%%%%%%%

\section{Conclusions}
\label{conclusions}

In conclusion, we have studied how the quantum vibrations of a suspended
carbon nanotube in the quantum dot regime affect the charge transport through it.
We have paid special attention to the respective signatures of the polaronic 
effect that arises from the position-dependent capacitive coupling 
of the nanotube with a nearby gate and the magneto-elastic coupling
in the presence of a transverse magnetic field. While both couplings act similarly in
the suppression of the current at the sequential tunneling order, their effect is qualitatively different
in the cotunneling regime. 

Our estimates for the polaron and magneto-elastic coupling constants show that 
these effects may be relevant to determine the current-voltage characteristics through 
suspended nanotubes. Thus,  charge transport could be efficiently used to demonstrate the 
quantum nature of the mechanical vibrations of these tiny objects.

\begin{acknowledgments}
This work has been supported by ANR through contract JCJC-036 NEMESIS 
and by the Nanosciences Foundation (Grenoble).
G.R. acknowledges support from the European networks SOLID and GEOMDISS, 
and from ANR contract QUANTJO.

\end{acknowledgments}

\appendix

\section{Rigorous derivation of \refE{eq:Hamiltonian3}}
\label{app:derivation_H1d}

In this Appendix, the effectively one-dimensional Hamiltonian \refe{eq:Hamiltonian3} 
for electrons in a suspended nanowire is derived and its conditions
of validity are obtained. 

For this, we start again with \refE{eq:Hamiltonian} and then
perform a unitary transformation ${\cal H}\rightarrow e^{iS}{\cal H}e^{-iS}$ where
$S=p_y u(x)-eB U(x)$, to obtain
\beqa
{\cal H}
&=&
\frac{[p_x-p_y\partial _x u(x)-eBy]^2+\bm{p}_\perp^2}{2m}+V(\bm{r})
\nonumber \\
&+&
\sum_n\left(
\frac{[P_n-p_yf_n(x)+eBF_n(x)]^2}{2M}+\frac{M\omega_n^2X_n^2}{2}
\right).
\nonumber \\
\label{eq:Hamiltonian4}
\eeqa
Here $\bm{p}_\perp=(p_y,p_z)$.
Assuming a strong confinement in the nanowire, we project
\refE{eq:Hamiltonian4} on the ground state
for the electron's transverse motion, and we obtain
\beqa
{\cal H}&\approx &\frac{[p_x-eB\langle y \rangle_0]^2}{2m}
+\eps_t +V_b(x)
\nonumber \\
&+& \sum_n\left(
\frac{[P_n+eBF_n(x)]^2}{2M}+\frac{M\omega_n^2X_n^2}{2}
\right.
\nonumber \\
&+&
\left.
\frac{eB\langle \{ y,p_y\} \rangle_0}{2m}X_nf_n'(x)
+
\frac{\langle p_y^2\rangle _0}{2M}f_n(x)^2
\right)
\nonumber\\
&+&
\frac{\langle p_y^2 \rangle_0}{2m}\left(\sum_nX_nf_n'(x)\right)^2
,
\label{eq:Hamiltonian4bis}
\eeqa
with
\beq
\eps_t=\eps_0+\frac{e^2B^2}{2m}\left[\langle y^2 \rangle_0-\langle y \rangle_0^2\right].
\label{eq:transverse-energy}
\eeq
In Eqs.~\refe{eq:Hamiltonian4bis}-\refe{eq:transverse-energy}, we denoted
$
\langle a \rangle_0=\int d\bm{r}_\perp\chi_0^* a
\chi_0
$
for an arbitrary function $a(\bm{r}_\perp,\bm{p}_\perp)$,
where $\bm{r}_\perp=(y,z)$ and $\chi_0$ is the 
ground-state wavefunction for transverse motion that solves
the Schr\"odinger equation:
\beq
\left[\frac{\bm{p}_\perp^2}{2m}+U_\mathrm{conf}(\bm{r}_\perp)\right]\chi_0(\bm{r}_\perp)=\eps_0\chi_0(\bm{r}_\perp).
\eeq
(Note that $\langle p_y \rangle_0=0$.)
The first term in \refE{eq:transverse-energy} is the ground-state energy $\eps_0$ for transverse motion,
the second term is a global diamagnetic shift.
Both will be absorbed into the chemical potential from now.

We now argue that the terms in the two last line of \refE{eq:Hamiltonian4}
can be safely neglected:

The first term would lead to a shift in the equilibrium position of the nanotube
that is characterized by a relative displacement
$\delta X_n/X_{n0}\sim (M/m) (X_{n0}/L)^2 \phi_n$
of the eigenmode $n$.
We also anticipated that the relevant magnetic field
dependance in the model enters through flux
$\sim B L X_{n0}$, see \refE{eq:flux}.

The second term (like the third one) acts even in the absence of magnetic field.
It generates an energy shift $\Delta \eps$ when the electron is
in the suspended section of the nanowire, with typical amplitude
$\Delta \eps/\omega_n\sim(X_{n0}/L_\perp)^2 (L/a)$. Here,
 $L_\perp$ is the transverse width of the wire and $a$ is a
cut-off at short distances (on the nanoscale) which regularizes the sum
$\sum_nf_n^2(x)=\delta(0)$ that diverges formally in the approximation of
continuous medium leading to \refE{eq:H-string}.
The parameter $a$ represents the microscopic atomic distance, the natural 
 length scale at which the continuous model ceases to be valid.

The last term leads to a renormalisation
of the oscillator's frequency when the electron is
in the suspended nanowire, with typical relative magnitude
$\delta\omega_n/\omega_n\sim (M/m).(X_{n0}/L)^2.(X_{n0}/L_\perp)^2$.
We can check that all terms give small correction in the case
of the single-wall carbon nanotube with parameters 
given in Sec. \ref{discussion}

Taking into account these simplifications in \refE{eq:Hamiltonian4} and making another
unitary transformation ${\cal H}\rightarrow e^{iS'} {\cal H} e^{-iS'}$, with $S'=eBx\langle y\rangle _0$,
we finally recover Equation \refe{eq:Hamiltonian3}.

\section{transition rates associated with cotunneling events}
\label{app:W_nm}

In this Appendix we obtain the transition rates associated with cotunneling events
following the method discussed in Ref. \onlinecite{koch:2006} 

In the absence of sequential tunneling term in \refE{Tmat}, the transition rate
\refe{golden} for the dot-phonon state to evolve from $(s,n)$ to $(s',m)$ while an electron
is transferred from lead $\nu$ to lead $\nu'$ is
\begin{widetext}
\beq
W^{(\nu\nu')}_{sn,s'm}
=
2\pi
\sum_{i_{\nu},f_{\nu}}
\sum_{i_{\nu'},f_{\nu'}} 
P^{(\mathrm{eq})}(i_{\nu})  P^{(\mathrm{eq})}(i_{\nu'}) 
 {\left| \left< f_{\nu},f_{\nu'},s',n \right|
{H}_T G_0 {H}_T
\left| i_{\nu},i_{\nu'},s,m \right> \right|}^2
 \delta \left( \Delta E_{\nu} + \Delta E_{\nu'} + \Delta_{ss'}  + \Delta_{mn} \right) 
\label{eqn:Wcot_gen}
\eeq
where $\Delta E_{\nu} = E_{f_{\nu}} -  E_{i_{\nu}}$,
$\Delta_{mn}=\omega [m-n]$, and
$\Delta_{ss'} =\eps_{ds}-\eps_{ds'}$.

The non-vanishing rates take the form
\begin{eqnarray}
W^{(\nu\nu')}_{0n,0m}
&=&
\frac{\Gamma_{\nu}\Gamma_{\nu'}}{2\pi}
\int d \xi
n_{\nu}(\xi)
[1- n_{\nu'}(\xi +\Delta_{nm})]  
\sum_{\sigma=\uparrow,\downarrow} {\left| \sum^{\infty}_{q=0}
\frac{M^{(\nu)*}_{nq} M^{(\nu')}_{mq}}
{\xi - \varepsilon_{d \sigma} + \Delta_{nq}+i\eta}
\right|}^2  \, .
\\
W^{(\nu\nu')}_{ \sigma n,\sigma 'm }
&=&
\frac{\Gamma_{\nu}\Gamma_{\nu'}}{2\pi}
\int d \xi
n_{\nu}(\xi) [1- n_{\nu'}(\xi +\Delta_{\sigma\sigma'}+\Delta_{nm})]  
 {\left| \sum^{\infty}_{q=0}
\frac{M^{(\nu)*}_{qm} M^{(\nu')}_{qn}}
{-\xi +\varepsilon_{d \sigma'} + \Delta_{mq}+i\eta}
\right|}^2  \, .
\end{eqnarray}

These expressions diverge as $\eta\rightarrow 0$.
A regularization scheme to remove this divergence while the sequential tunneling events are
properly taken was discussed in Ref. \onlinecite{koch:2006}. It consists in removing
an ${\cal O}(1/\eta)$-term, which corresponds to sequential tunneling events and is already accounted for in 
the rates \refe{eq:Wseqtunn}, while keeping the next order term [${\cal O}(1)$ term].
On the end, the regularized cotunneling transition rates read
\beqa
&&
W^{(\nu\nu')}_{0n,0m}  =
\frac{\Gamma_{\nu}\Gamma_{\nu'}}{2\pi}
\sum_{\sigma=\uparrow,\downarrow}
\sum^{\infty}_{p,q=0}
M^{(\nu)*}_{nq}  M^{(\nu')}_{mq}  M^{(\nu)}_{np}  M^{(\nu')*}_{mp}
\mathcal{F}
\left( \mu_{\nu} ; \mu_{\nu'}  - \Delta_{nm} ;
\varepsilon_{d \sigma} -\Delta_{nq};
\varepsilon_{d \sigma}- \Delta_{np}
\right)
\\
&&
W^{(\nu\nu')}_{ \sigma n,\sigma'm}
=
\frac{\Gamma_{\nu}\Gamma_{\nu'}}{2\pi}
\sum^{\infty}_{p,q=0}
M^{(\nu)*}_{qm}  M^{(\nu')}_{qn} M^{(\nu)}_{pm}  M^{(\nu')*}_{pn}
\mathcal{F}\left( \mu_{\nu} ; \mu_{\nu'}  -\Delta_{\sigma\sigma'}-\Delta_{nm};
\varepsilon_{d \sigma'} + \Delta_{mp};
\varepsilon_{d,\sigma'} + \Delta_{mq}
\right)
\nonumber
\\
\eeqa
where
\begin{eqnarray}
\mathcal{F}(E_1,E_2,\varepsilon_1,\varepsilon_2) 
&=& 
\lim_{\eta \rightarrow 0}
\mbox{Re}
\left[
\int dE
\frac{n(E-E_1)[1-n(E-E_2)]}{(E - \varepsilon_1 + i \eta)(E - \varepsilon_2 - i \eta)}
-
{\cal O}\left(\frac 1 \eta\right)
\right]
\nonumber \\
&=&
\frac{n_B(E_2-E_1)}{\varepsilon_1-\varepsilon_2}
\mbox{Re}
\left[
\Psi\left(\frac{1}{2} + i \frac{E_2 - \varepsilon_1}{2\pi T} \right)
-
\Psi\left(\frac{1}{2} + i \frac{E_2 - \varepsilon_2}{2\pi T} \right)
\right.
\nonumber \\
&&
\left.
-
\Psi\left(\frac{1}{2} + i \frac{E_1 - \varepsilon_1}{2\pi T} \right)
+
\Psi\left(\frac{1}{2} + i \frac{E_1 - \varepsilon_2}{2\pi T} \right)
\right],
\end{eqnarray}
$n_B(E)$ is the Bose distribution function, and $\Psi$ is the digamma function.
\end{widetext}

\bibliographystyle{apsrev4-1}

\bibliography{biblioNEMS}

\end{document}